\documentclass[aps,prl,floats,superscriptaddress,twocolumn,showpacs]{revtex4}%
\usepackage{epsfig}
\usepackage{dcolumn}
\usepackage{bm}
\usepackage{amsmath}
\usepackage{amsfonts}
\usepackage{amssymb}
\usepackage{graphicx}
\usepackage[latin1]{inputenc}%
\begin{document}
\title{Critical behavior at Mott-Anderson transition: a TMT-DMFT perspective}
\author{M. C. O. Aguiar}
\affiliation{Departamento de Física, Universidade Federal de Minas Gerais, Av. Antônio
Carlos, 6627, Belo Horizonte, MG, Brazil}
\author{V. Dobrosavljevi\'{c}}
\affiliation{Department of Physics and National High Magnetic Field Laboratory, Florida
State University, Tallahassee, FL 32306, USA}
\author{E. Abrahams}
\affiliation{Center for Materials Theory, Serin Physics Laboratory, Rutgers University, 136
Frelinghuysen Road, Piscataway, New Jersey 08854, USA}
\author{G. Kotliar}
\affiliation{Center for Materials Theory, Serin Physics Laboratory, Rutgers University, 136
Frelinghuysen Road, Piscataway, New Jersey 08854, USA}


\begin{abstract}
We present a detailed analysis of the critical behavior close to the
Mott-Anderson transition. Our findings are based on a combination of numerical
and analytical results obtained within the framework of Typical-Medium
Theory (TMT-DMFT) - the simplest extension of dynamical mean field theory
(DMFT) capable of incorporating Anderson localization effects. 
By making use of previous scaling studies of Anderson impurity models close to 
the metal-insulator transition, we solve this problem analytically and reveal 
the dependence of the critical behavior on the particle-hole symmetry.
Our main result is that, for sufficiently strong disorder, the Mott-Anderson 
transition is characterized by a precisely defined two-fluid behavior, in which 
only a fraction of the electrons undergo a ``site selective'' Mott localization; 
the rest become Anderson-localized quasiparticles.

\end{abstract}
\maketitle

Many strongly correlated materials find themselves close to Mott localization
\cite{mott74} \ - a process through which all the valence electrons within a
narrow band turn into localized magnetic moments. 
In real systems, disorder introduced by doping or impurities often cannot be
neglected, as it provides an alternative fundamental mechanism for suppressing
metalicity, through the
process of Anderson localization \cite{anderson58}. The effects of weak
interactions in this regime have been studied using perturbative methods
~\cite{lr}, but these approaches cannot describe the strong correlation
effects associated with incipient magnetism and Mott localization.

Which of these two routes to localization - Anderson or Mott - dominate? In most 
cases, simple estimates show that both effects play a comparable role and both 
need to be taken into account. Most existing theories are not able to combine these 
two fundamental processes in the same framework, and this conceptual difficulty
has provided the essential pitfall in our understanding of the metal-insulator
transition (MIT).

At the moment, the most successful theory for the Mott transition is based on
dynamical mean field theory (DMFT) \cite{dmftrev} ideas. By replacing the
environment of each site by its average value, the original version of this
theory proved unable to describe the spatial fluctuation effects associated
with the approach to the Anderson transition. Very recent work \cite{tmt},
however, identified the conceptually simplest extension of DMFT capable to
overcome these shortcomings - the Typical-Medium Theory (TMT-DMFT). In 
the non-interacting limit, this theory provides a reasonable picture of the 
Anderson transition, as established by quantitative comparison \cite{tmt} with 
exact (numerical) results.

The TMT-DMFT method was first applied to the disordered Hubbard model by
Byczuk \textit{et al.}~\cite{vollhardt}, who obtained the phase diagram for
this problem from the numerical solution using the Numerical Renormalization
Group (NRG) method for the impurity solver. However, the physical nature of the 
phases and of the phase transition was not investigated in that 
numerical study.

The task of elucidating the physical mechanism and the precise form of the
Mott-Anderson critical point within the TMT-DMFT description is the main subject 
of this Letter. By making use of previous scaling studies \cite{sces07}
of Anderson impurity models close to the MIT, we present a detailed analytic
solution for this problem, which emphasizes the dependence of the system
properties on its particle-hole symmetry.
Our main finding is that, for sufficiently strong disorder, the
physical mechanism behind the Mott-Anderson transition is the formation of 
two fluids, a behavior that is surprisingly reminiscent of
the phenomenology proposed for doped semiconductors \cite{paalanenetal88}.
Here, only a fraction of the electrons (sites) undergo Mott localization; the
rest can be described as Anderson-localized quasiparticles. Thus, in our 
picture the Mott-Anderson transition can be seen as reminiscent of the 
``orbitally selective'' Mott localization~\cite{osmt}; precisely, here
we have a ``site selective'' Mott transition, since it emerges in a 
spatially resolved fashion.

\textit{TMT-DMFT and order parameters }- We consider a half-filled Hubbard
model~\cite{dmftrev} with random site energies, as given by the Hamiltonian%
\begin{equation}
H=-V\sum_{<ij>\sigma}c_{i\sigma}^{\dagger}c_{j\sigma}+\sum_{i\sigma
}\varepsilon_{i}n_{i\sigma}+U\sum_{i}n_{i\uparrow}n_{i\downarrow}.
\label{hamiltonian}%
\end{equation}
Here, $c_{i\sigma}^{\dagger}$ ($c_{i\sigma}$) creates (destroys) a conduction
electron with spin $\sigma$ on site $i$, $n_{i\sigma}=c_{i\sigma}^{\dagger
}c_{i\sigma}$, $V$ is the hopping amplitude, and $U$ is the on-site repulsion.
The random on-site energies $\varepsilon_{i}$ follow a distribution
$P(\varepsilon)$, which is assumed to be uniform and have width $W$.

TMT-DMFT \cite{tmt,vollhardt} maps the lattice problem onto an ensemble of
single-impurity problems, corresponding to sites with different values of the
local energy $\varepsilon_{i}$, each being embedded in a typical
effective medium which is self-consistently calculated. In contrast to
standard DMFT \cite{screening}, TMT-DMFT determines this effective medium by
replacing the spectrum of the environment (``cavity'') for each site by its
typical value, which is determined by the process of \textit{geometric} averaging. 
For a simple semi-circular model density of states, the corresponding bath
function is given by \cite{tmt,vollhardt} $\Delta(\omega)=V^{2}G_{typ}%
(\omega)$, with $G_{typ}(\omega)=\int_{-\infty}^{\infty}d\omega^{\prime}%
\rho_{typ}(\omega^{\prime})/(\omega-\omega^{\prime})$ being the Hilbert
transform of the geometrically-averaged (typical) local density of states
(LDOS) $\rho_{typ}(\omega)=\exp\{\int d\varepsilon P(\varepsilon)\ln
\rho(\omega,\varepsilon)\}$. Given the bath function $\Delta(\omega)$, one
first needs to solve the local impurity models and compute the local spectra
$\rho(\omega,\varepsilon)=-\pi^{-1}\operatorname{Im}G(\omega,\varepsilon)$,
and the self-consistency loop is then closed by the the geometric averaging procedure.

To qualitatively understand the nature of the critical behavior, it is useful
to concentrate on the low-energy form for the local Green's functions, which
can be specified in terms of two Fermi liquid parameters as
\begin{equation}
G(\omega,\varepsilon_{i})=\frac{Z_{i}}{\omega-\tilde{\varepsilon}_{i}%
-Z_{i}\Delta(\omega)},
\end{equation}
where $Z_{i}$ is the local quasi-particle (QP) weight and $\tilde{\varepsilon
}_{i}$ is the renormalized site energy \cite{screening}. The parameters
$Z_{i}$ and $\tilde{\varepsilon}_{i}$ can be obtained using any quantum
impurity solver, but to gain analytical insight here we focus on the
variational calculation provided by the 
``four-boson'' technique (SB4) of Kotliar and
Ruckenstein~\cite{sb4}, which is known to be quantitatively accurate at $T=0$.
We should stress, though, that most of our analytical results rely only on
Fermi liquid theorems constraining the qualitative behavior at low energy, and
thus do not suffer from possible limitations of the SB4 method.

Within this formulation, the metal is identified by nonzero QP weights $Z_{i}$
on \textit{all} sites and, in addition, a nonzero value for both the
typical and the average [$\rho_{av}(\omega)=\int d\varepsilon P(\varepsilon
)\rho(\omega,\varepsilon)$] LDOS. Mott localization (i.e. local moment
formation) is signaled by $Z_{i}\longrightarrow0$ \cite{screening}, while
Anderson localization corresponds to $Z_{i}\neq0$ and
$\rho_{av}\neq0$, but $\rho_{typ}=0$ \cite{anderson58,tmt}. While Ref.
\cite{vollhardt} concentrated on $\rho_{typ}$ and $\rho_{av}$, we find it
useful to simultaneously examine the QP weights $Z_{i}$, in order to provide a
complete and precise description of the critical behavior.

\begin{figure}[t]
\begin{center}
\vspace{0.6cm} \includegraphics[
trim=0.0in 0.0in 0.0in 0in,
height=1.9in, width=2.5in ]
{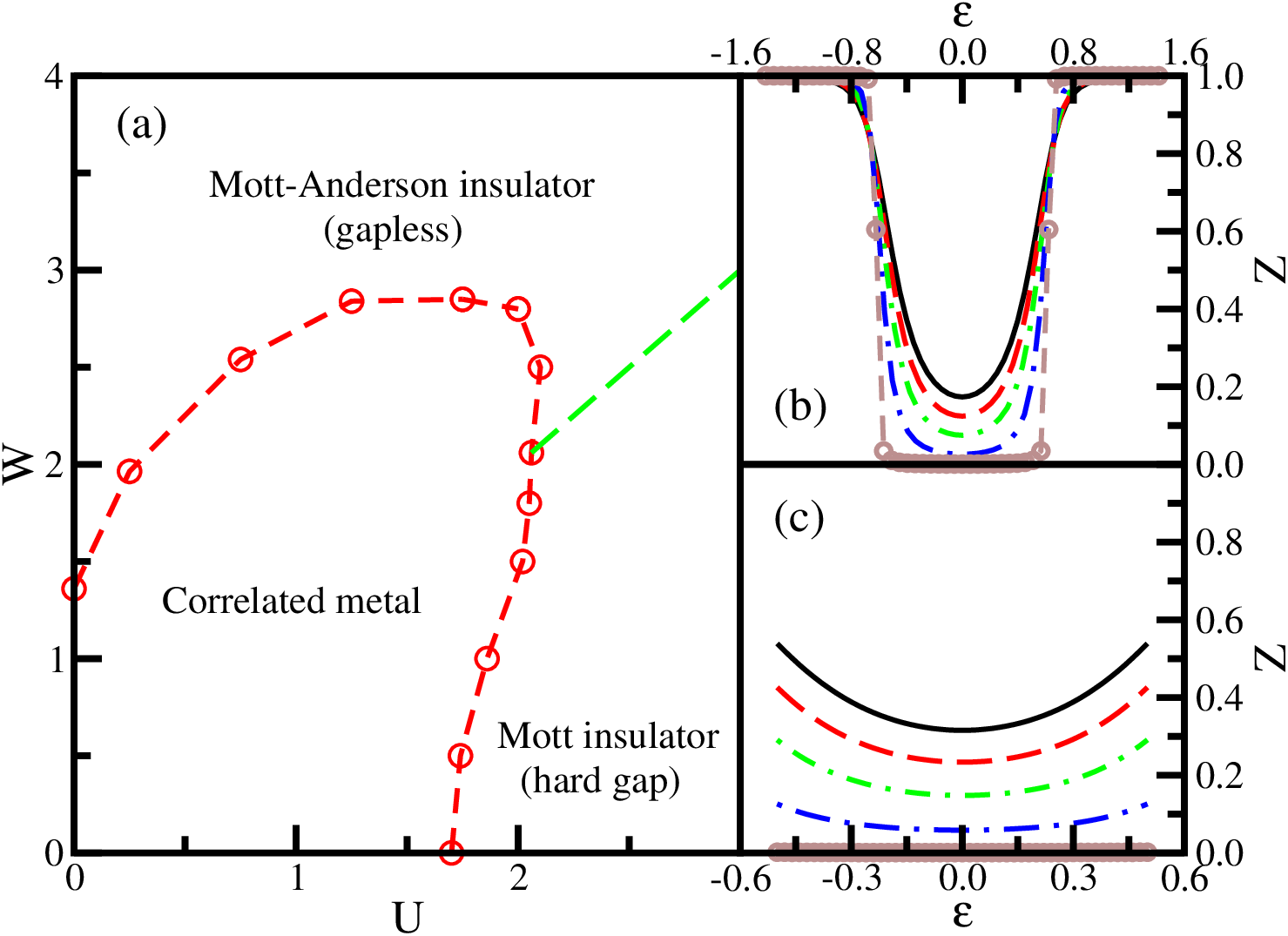}
\end{center}
\caption{(Color online) (a) $T=0$ phase diagram for the disordered half filled
Hubbard model, obtained from the numerical SB4 solution of TMT-DMFT. Panels
(b) and (c) show the evolution of the quasi-particle weight $Z(\varepsilon
_{i})$ in the critical region. Behavior at (b) the Mott-Anderson transition
($W>U$) is illustrated by increasing disorder $W$ $=2.5$, $2.6$, $2.7$, $2.8$,
$2.83$ (from the black curve to the brown one), for fixed $U=1.25$; and at 
(c) the Mott-like transition ($W<U$) by
increasing the interaction interaction $U=1.5$, $1.6$, $1.7$, $1.8$ and
$1.86$, at fixed disorder $W=1.0$. }%
\end{figure}

\textit{Phase diagram }- Using our SB4 method, the TMT-DMFT equations can be
numerically solved to very high accuracy, allowing very precise
characterization of the critical behavior. In presenting all numerical results
we use units such that the bandwidth $B=4V=1$. Fig.~1a shows the resulting
$T=0$ phase diagram at half filling, which generally~\cite{phdiagram}
agrees with that of
Ref.~\cite{vollhardt}. By concentrating first on the critical behavior of the
QP weights $Z_{i}$, we are able to clearly and precisely distinguish the metal
from the insulator. We find that at least some of the $Z_{i}$ vanish all along
the phase boundary.
By taking a closer look, however, we can distinguish two types of critical 
behavior, as follows.

\textit{Mott-Anderson vs. Mott-like transition -} For sufficiently strong
disorder ($W>U$), the Mott-Anderson transition proves qualitatively different
than the clean Mott transition, as seen by examining the critical behavior of
the QP weights $Z_{i}=Z(\varepsilon_{i})$ (Fig.~1b). Here $Z_{i}\rightarrow0$
only for $0<|\varepsilon_{i}|<U/2$ , indicating that only \textit{a fraction}
of the electrons turn into localized magnetic moments. The rest show
$Z_{i}\rightarrow1$ and undergo Anderson localization (see below). Physically,
this regime corresponds to a spatially inhomogeneous system, with
Mott fluid droplets interlaced with
regions containing Anderson-localized quasiparticles. In contrast, for weaker
disorder ($W<U$) the transition retains the conventional Mott character. In
this regime $Z_{i}\rightarrow0$ on all sites (Fig.~1c), corresponding to Mott
localization of all electrons.

\begin{figure}[ptb]
\vspace{0.3cm}
\par
\begin{center}
\includegraphics[
trim=0.0in 0.0in 0.0in 0in,
height=1.9in, width=2.4in ]
{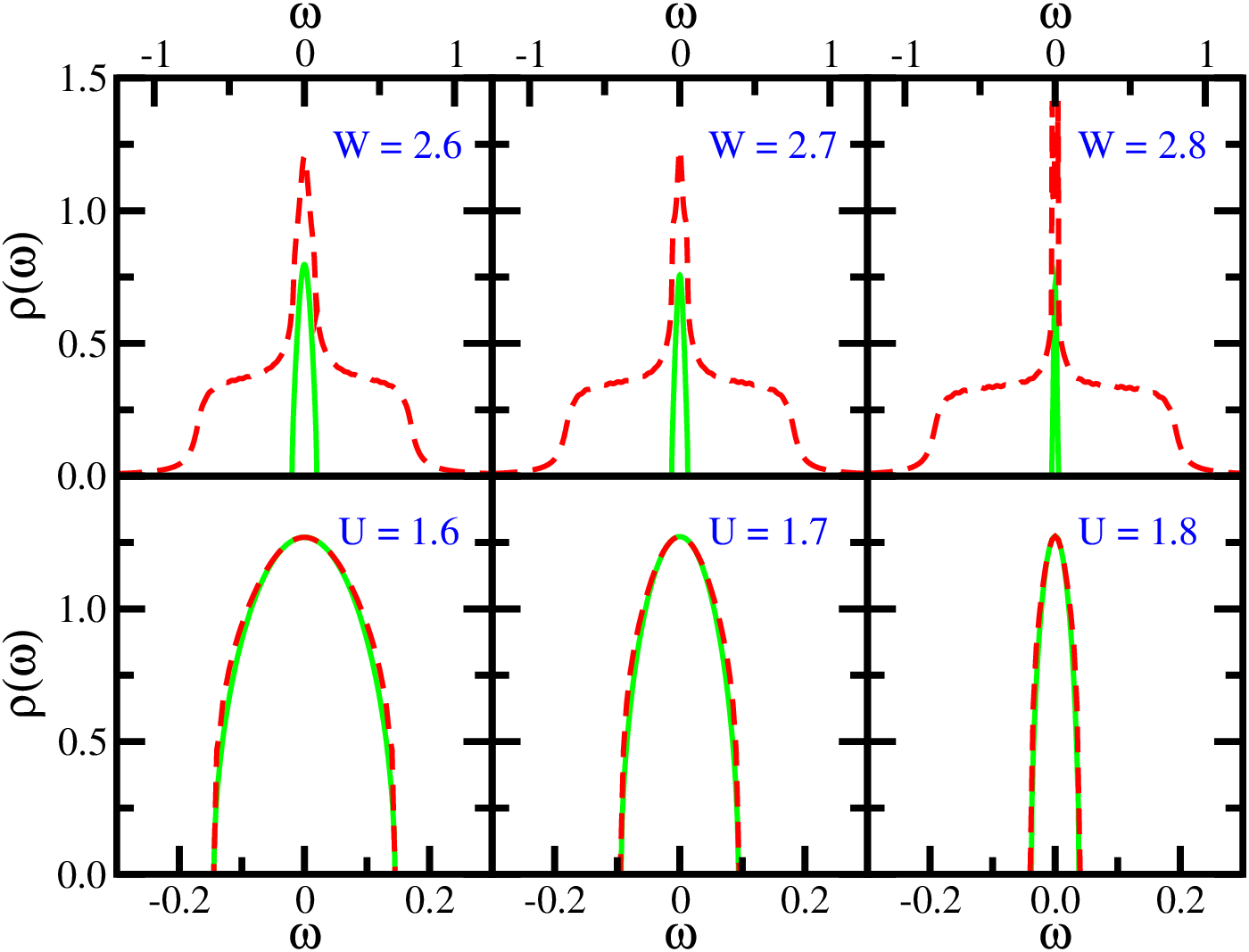}
\end{center}
\caption{(Color online) Frequency dependence of $\rho_{typ}$ (full line) and
$\rho_{av}$ (dashed line) in the critical region. Results in top panels
illustrate the approach to the Mott-Anderson transition ($W>U$) at $U=1.25$;
the bottom panels correspond to the Mott-like transition ($W<U$) at $W=1.0$. 
}%
\end{figure}

\textit{Wavefunction localization} - To more precisely characterize the
spatial fluctuations of the quasiparticle wavefunctions, we compare the
behavior of the typical ($\rho_{typ}$) and the average ($\rho_{av}$) LDOS.
The approach to the Mott-Anderson transition ($W>U$) is
illustrated by increasing disorder $W$ for fixed $U=1.25$ (Fig.~2 - top
panels). Only those states within a narrow energy range ($\omega<t$, see also
Fig.~4) around the band center (the Fermi energy) remain spatially delocalized
($\rho_{typ}\sim\rho_{av}$), due to strong disorder screening
\cite{screening,sces07} within the Mott fluid (sites showing $Z_{i}%
\rightarrow0$ at the transition). The electronic states away from the band 
center (i.e. in the band tails) quickly get Anderson-localized, displaying
large spatial fluctuations of the wavefunction amplitudes \cite{tmt} and
having $\rho_{typ}\ll\rho_{av}$. 

The spectral weight of the delocalized states (states in the range $\omega<t$) 
decreases with disorder and
vanishes at the transition, indicating the Mott localization of this fraction 
of electrons. At this critical point, the crossover scale $t$ also vanishes. In 
contrast, the \textit{height} $\rho_{typ}(0)$ remains finite at the transition, 
albeit at a reduced $W$-dependent value, as compared to the clean limit. More
precise evolution of $\rho_{typ}(0)$ is shown in Fig.~3a, demonstrating its
critical jump.

Behavior at the Mott-like transition ($W<U$) is dramatically different (Fig.~2
- bottom panel). Here $\rho_{typ}\approx\rho_{av}$ over the entire QP band,
indicating the absence of Anderson localization. It proves essentially
identical as that established for the disordered Hubbard model within standard
DMFT~\cite{screening}, reflecting strong correlation-enhanced screening of
disorder \cite{sces07}, where both $\rho_{av}(\omega=0)$ and $\rho_{typ}(\omega=0)$ 
approach the bare ($W=0$) value (see also Fig.~3b). Similar results were found in 
Ref. \cite{vollhardt}, but an explanation was not provided. 

The corresponding 
pinning \cite{screening,sces07} for $\rho(\omega=0,\varepsilon)$ is shown 
in the insets of Fig.~3, both for the Mott-Anderson and the Mott-like transition. 
In the 
Mott-Anderson case, this mechanism applies only within the Mott fluid 
($|\varepsilon|<U/2$), while within the Anderson fluid 
($|\varepsilon|>U/2$) it assumes smaller values, explaining the 
reduction of $\rho_{typ}(0)$ in this case.
We suggest that this spatial distribution of the DOS at the Fermi energy
(each $\varepsilon$ corresponds to a different position in the lattice)
could be probed by scanning tunneling microscopy experiments.

\begin{figure}[t]
\begin{center}
\includegraphics[
trim=0.0in 0.0in 0.0in 0in,
height=1.9in, width=2.5in ]
{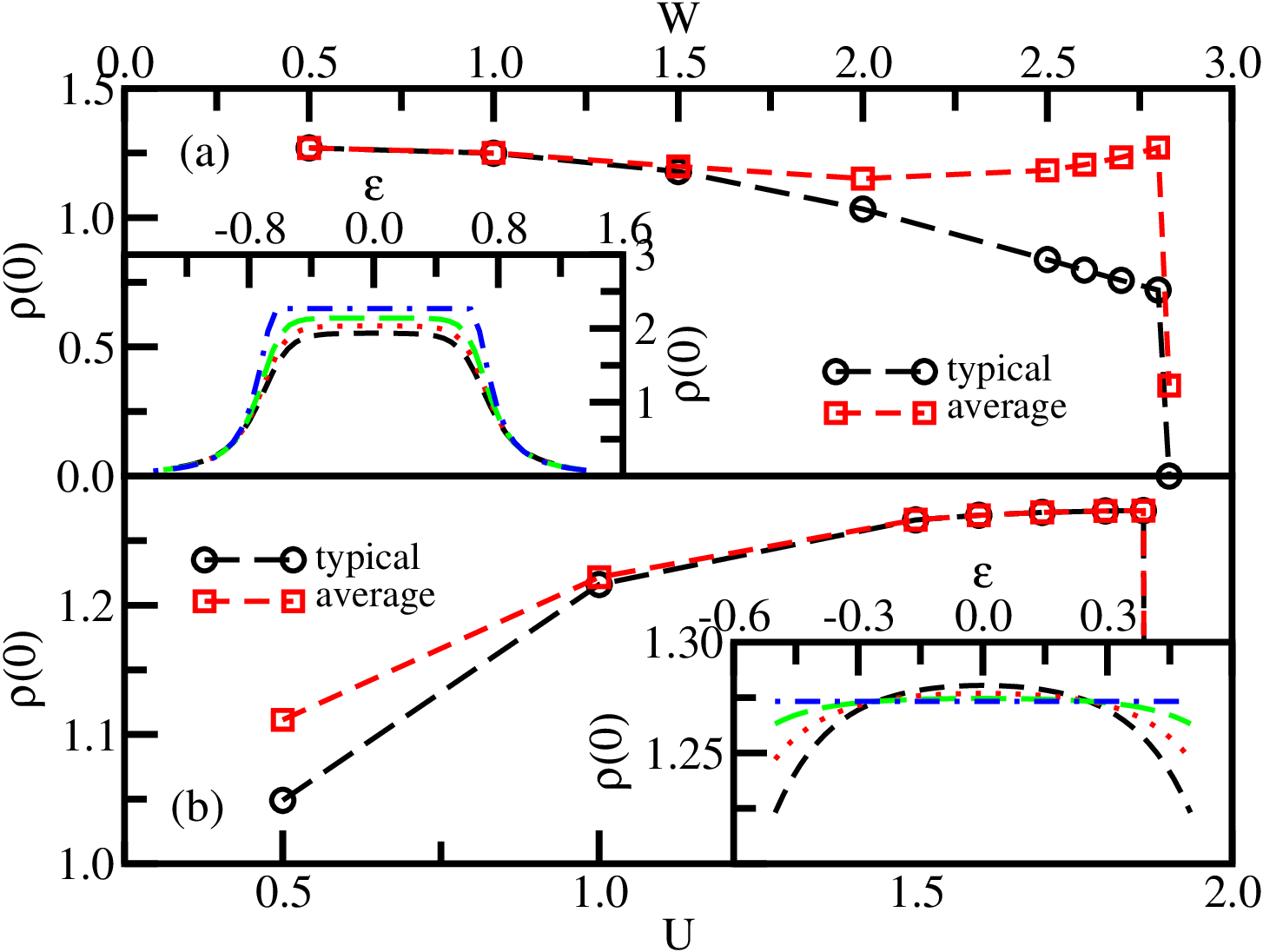}
\end{center}
\caption{(Color online) Typical and average values of $\rho(0)$ as the
metal-insulator transition is approached for (a)~$U=1.25$ and (b) $W=1.0$. The
insets show $\rho(0)$ as a function of $\varepsilon$ for (a) $W=2.5$, $2.6$,
$2.7$ and $2.83$ (from the black curve to the blue one) and (b) $U=1.5$,
$1.6$, $1.7$ and $1.86$.}%
\end{figure}

\textit{Analytical solution} -
Within our SB4 approach, the TMT-DMFT order-parameter function $\rho
_{typ}(\omega)$ satisfies the following self-consistency condition%
\begin{align}
\rho_{typ}(\omega) &  =\exp \int d\varepsilon P\left(  \varepsilon\right)
\left\{  \ln[V^{2}Z^{2}(\varepsilon)\rho_{typ}(\omega)] \right. \nonumber\\
&  -\ln[(\omega-\widetilde{\varepsilon}(\varepsilon)-V^{2}Z(\varepsilon
)\operatorname{Re}G_{typ}(\omega))^{2}\nonumber\\
&  \left. +(\pi V^{2}Z(\varepsilon)\rho_{typ}(\omega))^{2}]\right\}.
\end{align}
While the solution of this equation is in general difficult, it
simplifies in the critical region, where the QP parameter functions
$Z(\varepsilon)$ and $\widetilde{\varepsilon}(\varepsilon)$ assume scaling
forms which we carefully studied in previous work \cite{sces07}. This
simplification allows, in principle, to obtain a closed solution for all
quantities. In particular, the crossover scale $t$, which defines the 
$\rho _{typ}(\omega)$ mobility edge (see Fig.~4 and Ref.~\cite{sces07}), is 
determined by setting $\rho_{typ}(\omega=t)=0$. 

Using this approach we obtain that, in the case of Mott-like transition
($W<U$), the critical behavior of all quantities reduces to that found in
standard DMFT~\cite{screening}, including $t\sim U_{c}(W)-U$ (in agreement
with the numerical results of Fig.~4b), perfect screening of site 
randomness~\cite{screening,sces07}, and the approach of $\rho_{av}(\omega=0)$ 
and $\rho_{typ}(\omega=0)$ to the clean value. The precise form of the critical 
behavior for the crossover scale $t$ is more complicated for the Mott-Anderson 
transition ($W>U$) (as confirmed by our numerical results in Fig.~4a), 
and this will not be discussed here.

\begin{figure}
\begin{center}
\includegraphics[
trim=0.0in 0.0in 0.0in 0in,
height=1.9in, width=2.5in ]
{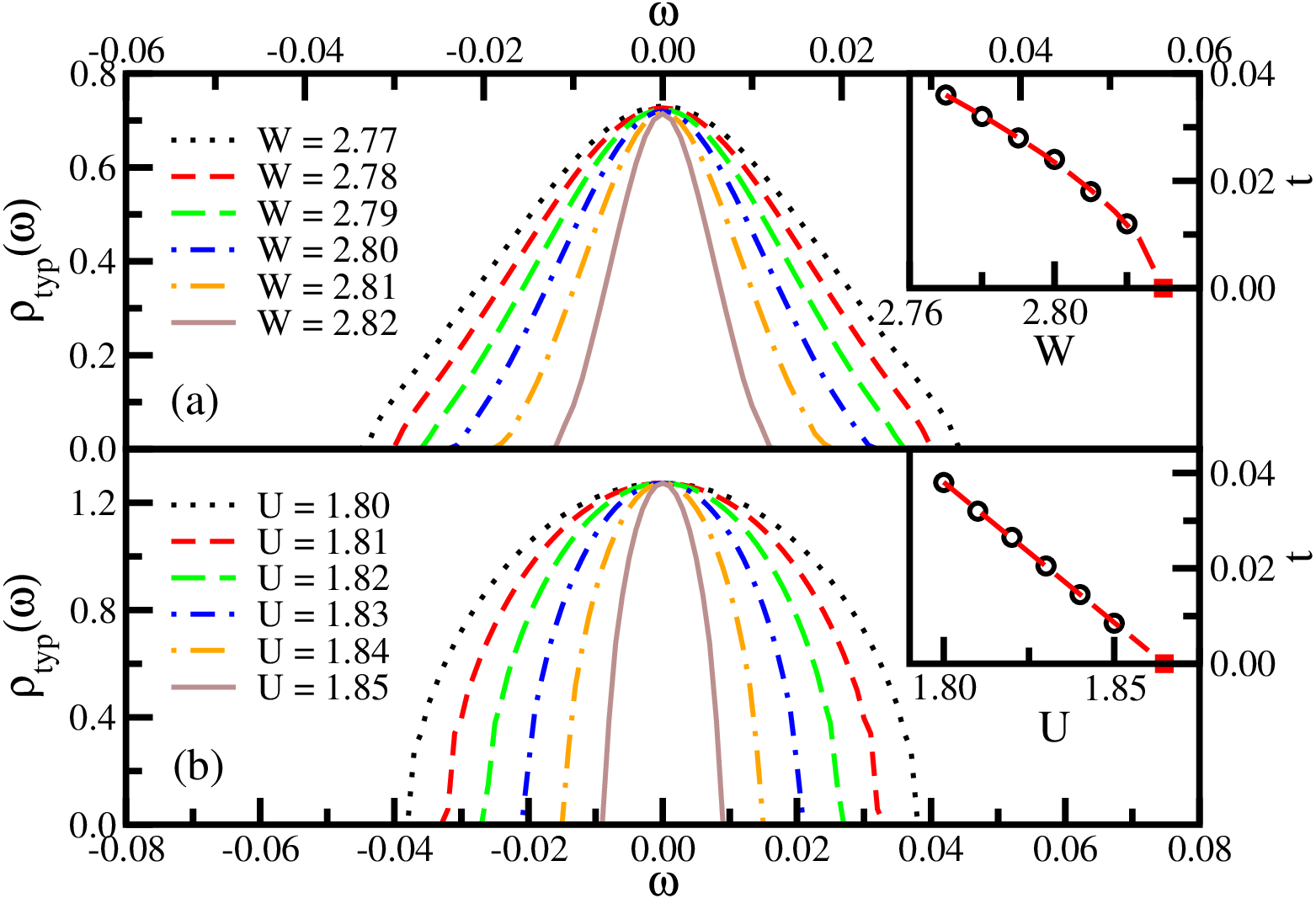}
\end{center}
\caption{(Color online) Frequency dependence of the typical DOS very close to
the metal-insulator transition for (a) the Mott-Anderson transition ($W>U$)
at $U = 1.25$ and (b) the Mott-like transition ($W<U$) at $W=1.0$. The insets 
show how, in both cases, the $\rho
_{typ}(\omega)$ bandwidth $t\rightarrow0$ at the transitions.}%
\end{figure}

Instead, we focus on elucidating the origin of the puzzling behavior of
$\rho_{c}=\rho_{typ}(\omega=0)$, which is known \cite{tmt} to vanish
linearly $\rho_{c}\sim(W_{c}-W)$ for $U=0$, but which we numerically find to
display a jump (i.e. a finite value) at criticality, as soon as interactions
are turned on. For $\omega=0$ our self-consistency condition reduces
\cite{PHsymmetry} to
\begin{equation}
\int d\varepsilon P\left(  \varepsilon\right)  \ln\frac{V^{2}Z^{2}%
(\varepsilon)}{\widetilde{\varepsilon}(\varepsilon)^{2}+\pi^{2}V^{4}%
Z^{2}(\varepsilon)\rho_{c}^{2}}=0,\label{omega0}%
\end{equation}
which further simplifies as we approach the critical point. Here, the QP
parameters $Z(\varepsilon)\longrightarrow0$ and $\widetilde{\varepsilon
}(\varepsilon)\sim$ $Z^{2}(\varepsilon)\ll$ $Z(\varepsilon)$ for the
Mott fluid ($|\varepsilon|<U/2$), while
$Z(\varepsilon)\longrightarrow1$ and $|\widetilde{\varepsilon}(\varepsilon
)|\longrightarrow$ $|\varepsilon-U/2|$ for the Anderson
fluid ($|\varepsilon|>U/2$), and we can write%
\begin{align}
0  & =\int_{0}^{U/2}\hspace{-12pt}d\varepsilon P\left(  \varepsilon\right)
\ln\frac{1}{\left(  \pi V\rho_{c}\right)  ^{2}}\nonumber\\
& -\int_{0}^{(W-U)/2}\hspace{-32pt}d\varepsilon P\left(  \varepsilon\right)
\ln[\left(  \varepsilon/V\right)  ^{2}+\left(  \pi V\rho_{c}\right)
^{2}].\label{rhoc}%
\end{align}

This expression becomes even simpler in the $U<<W$ limit, giving
\begin{equation}
\frac{U}{W}\ln\frac{1}{\pi V\rho_{c}}+a-bV\rho_{c}+O[\rho_{c}^{2}%
]=0,\label{smallU}%
\end{equation}
where $a(W,U)=(1-U/W)\{1-\ln[(W-U)/2V]\}$ and $b=\frac{2\pi^{2}V}{W}$. This
result reproduces the known result \cite{tmt} $\rho_{c}\sim(W_{c}-W)$ at
$U=0$, but dramatically different behavior is found as soon as $U>0$. Here,
a \textit{non-analytic} (singular) contribution emerges from the
Mott fluid ($|\varepsilon|<U/2$), which
assures that $\rho_{c}$ must remain finite at the critical point, consistent
with our numerical results (see Fig.~3). Note that the second term in 
Eq.~(\ref{rhoc}), coming from the Anderson
fluid ($|\varepsilon|>U/2$), vanishes in the case of a
Mott-like transition ($U>W$), and our result reproduces the standard
condition $\pi\rho_{c}V=1$~\cite{screening}, which corresponds to the
clean limit. 

A further glimpse on how the condition $\pi\rho_{c}V=1$ is gradually violated 
as we cross on
the Mott-Anderson side is provided by solving Eq.~(\ref{rhoc}) for $U\precsim
W$ limit, giving
\begin{equation}
\rho_{c}\approx\frac{1}{\pi V}\left[  1-\frac{1}{24}\left(  \frac{W}%
{V}\right)  ^{2}\left(  1-\frac{U}{W}\right)  ^{3}\right]  ,
\end{equation}
again consistent with our numerical solution \cite{largeU}.

But what is the physical origin of the jump in $\rho_c$? To see it, note that 
the singular form of the first term in Eq.~(\ref{rhoc}) comes from the
Kondo pinning \cite{screening} $\widetilde{\varepsilon}(\varepsilon)\sim$ 
$Z^{2}(\varepsilon)\ll$ $Z(\varepsilon)$ within the Mott fluid. 
This behavior reflects the particle-hole
symmetry of our (geometrically averaged) $\rho_{typ}(\omega=0)$ bath function,
which neglects site-to-site cavity fluctuations present, for example, in more
accurate statDMFT theories \cite{statdmft}. Indeed, in absence of
particle-hole symmetry, one expects \cite{screening} $\widetilde{\varepsilon
}(\varepsilon)\sim$ $Z(\varepsilon)$, and the resulting $\varepsilon
$-dependence should cut-off the log singularity responsible for the jump in
$\rho_{c}$. This observation provides a direct path to further refine the
TMT-DMFT approach, reconciling the present results with previous
statDMFT findings \cite{statdmft}. As a next step, one should apply
the TMT ideas to appropriately chosen effective
models~\cite{effmodel}, in order to eliminate those features
reflecting the unrealistic particle-hole symmetry built in the current theory.
We emphasize that the two-fluid picture is a consequence of only a  
fraction of the sites showing $Z \rightarrow 0$ and is not dependent on either  
particle-hole symmetry or the consequent jump in the DOS.

\textit{Conclusions} - \label{conclusion} This Letter explores the TMT-DMFT
critical region of the Mott-Anderson transition. We show how key insight can
be obtained by focusing on the evolution of the local quasiparticle weights
$Z_{i}$ as a \textit{second order parameter} describing tendency to Mott
localization, in addition to the Anderson-like TMT order parameter $\rho
_{typ}$. This analysis reveals the fundamental two-fluid character of the
Mott-Anderson transition, consistent with the phenomenology proposed for 
doped semiconductors~\cite{paalanenetal88}. Physically, it
describes spatially inhomogeneous situations, where the Fermi liquid
quasiparticles are destroyed only in certain regions - the
Mott droplets - but remain coherent elsewhere.
Understanding the details of such ``site selective'' Mott
transitions
should be viewed as an indispensable first step in solving the long-standing
problem of metal-insulator transitions in disordered correlated systems.

We thank Dieter Vollhardt for useful discussions. This work was supported by
PRPq/UFMG and CNPq grant 473987/2007-4 (M.C.O.A.) and NSF grants DMR-0234215
and DMR-0542026 (V.D.), and DMR-0096462 (G.K.).


\begin{thebibliography}{99}
\bibitem {phdiagram} We do not discuss the coexistence region found in
Ref.~\cite{vollhardt}, because we focus 
on criticality within the metallic phase. We do not find any ``crossover'' regime such 
as reported in Ref.~\cite{vollhardt}, the existence of which seems inconsistent with 
the generally sharp distinction between a metal and an insulator at $T=0$.
\bibitem {PHsymmetry}For our model $\mbox{Re}G_{typ}(0)=0$ by particle-hole symmetry.
\bibitem {largeU} In this respect, our results differ from those in Ref.~\cite{vollhardt}, 
where $\rho_c$ is shown to vanish for large $U$. This seems to occur within what the 
authors call the ``crossover'' regime, which is not present in our phase diagram (see 
Ref.~\cite{phdiagram}).
\end{thebibliography}
\end{document}